# Étude des dimensions spécifiques du contexte dans un système de filtrage d'informations

Djallel Bouneffouf

Department of Computer Science, Télécom SudParis, UMR CNRS Samovar, 91011 Evry Cedex, France

`Djallel.Bouneffouf@it-sudparis.eu`

**Abstract.** Que ce soit dans le contexte des systèmes d'information d'entreprise, du commerce électronique, de l'accès au savoir et aux connaissances ou même des loisirs, la pertinence de l'information délivrée aux usages constitue un fact eur clé du succès ou du rejet de ces systèmes d'information.
La qualité d'accès est conditionnée donc par l'accès à la bonne information, au bon moment, à l'endroit et sur le support choisi. Dans cette optique, il est important de tenir compte des besoins et des connaissances des utilisateurs lors de l'accès à l'information ainsi que de la situation contextuelle afin de lui fournir une information pertinente, adaptée à ses besoins et son contexte d'utilisation. Dans ce qui suit nous décrivons le prélude d'un projet qui essaye de regrouper tous ces besoins pour améliorer les systèmes d'information.

## 1      Introduction

La surabondance de l'information dans les systèmes d'information a engendré la dégradation de la qualité des résultats retournés à l'utilisateur (Dervin et al., 86), (Shamber, 94).

Les Smartphones ou autres medias internet mobile envahissent les poches des cadres des sociétés et commencent à être utilisés à des fins professionnelles. Ce nouveau cadre d'utilisation accentue le besoin et la nécessité de prendre en considération des informations du contexte pour améliorer la remontée des informations pertinentes aux utilisateurs. En effet, vu les contraintes et particularités techniques des appareils mobiles (zone d'affichage limitée, ...), nous assistons à une pratique de recherche différente de celle pratiquée au bureau.

Des études sur les mobinautes (Kamvar et al., 07) montrent que la majorité des utilisateurs ne consultent que la première page des résultats. De plus, d'après les études dans « Sohn et al., 08 », 72 % des besoins informationnels des utilisateurs mobiles sont liés à des facteurs contextuels notamment la localisation et le temps. En clair, le problème n'est pas tant la disponibilité de l'information, mais sa pertinence relative-ment à un contexte d'utilisation spécifique. C'est pourquoi les travaux en recherche contextuelle d'information (RCI) ont vu le jour ces dernières années (Ingwersen et al., 05), dans le but d'optimiser la pertinence des résultats de recherche.

Les recherches actuelles en RCI se sont focalisées sur la modélisation et l'exploitation de 5 dimensions spécifiques du contexte :
- la dimension dispositif (Goker et al, 02),
- la dimension tâche/problème (Jansen et al., 07),
- la dimension contexte du document (Xie, 08),
- la dimension spatio-temporelle (Tao et al., 03),
- la dimension contexte utilisateur (Timothy et al., 05)

Nous remarquons que la recherche en terme de contexte et de mobilité était surtout abordé au niveau recherche d'information mais, peu ou pas du tout entamé du coté du filtrage d'information, de ce fait nous nous proposons d'étudier les différentes dimensions du contexte et les appliquer au filtrage d'information dans un cadre de mobilité.

## 2  PROBLEMATIQUE

Le courant dans lequel s'inscrira ce projet est celui des « filtrage d'information contextuelle». Concrètement, l'objectif sera de développer une démarche de filtrage d'information adaptée au contexte et à la mobilité.

Lors de ce travail il faudra : décrire les modèles de données utiles dans le filtrage, mettre en place des méthodes sur des sous-problèmes liés à la masse d'information contenue dans le Système d'Information, le temps de réponse et l'accès aux informations sécurisées.

L'étude consiste donc à développer un cadre à la fois théorique et logiciel permettant l'identification des actions de l'utilisateur, la synthèse du contexte de ces actions, le filtrage des actions pertinentes, dans le but de proposer des méthodes qui amélioreraient la pertinence des informations remontées aux utilisateurs.

Tout au long des itérations de développement et d'expérimentation de cette plate-forme, les problématiques de recherche suivantes seront abordées :
• Comment utiliser les dimensions spécifiques du contexte venu du monde de la RI en vue d'obtenir des indices de pertinences fiables pour le filtrage d'information ?
. Proposer une méthode de calcul et de combinaison des scores (géographique et temporel, etc......) de pertinence des documents.
• Comment assurer l'acceptation de cette plateforme de modélisation de profil malgré les craintes liées à la confidentialité d'information ?
La démarche développée dans le cadre de ce projet trouvera plusieurs utilisations parmi lesquelles :
- L'élaboration d'un modèle de profil générique pour un système d'information mobile.
- L'élaboration du premier Système d'Information mobile contenant un système de filtrage d'information.

# 3    ETAT DE L'ART

## A)    FILTRAGE D'INFORMATION

Depuis que le nombre d'utilisateurs Internet et la disponibilité de l'information sous une forme électronique ont exponentiellement augmenté, le développement d'outils qui manipulent ce flux important d'informations est devenu nécessaire.

Les systèmes de recherche d'information ont apporté une aide précieuse dans la recherche des informations pertinentes, mais la quantité d'informations qu'ils génèrent en réponse à une requête donnée est très considérable. Ceci, contraint les utilisateurs à dépenser plus de temps pour rechercher l'information exacte.

Pour filtrer ces informations, c'est-à-dire sélectionner seulement celles qui nous intéressent, les systèmes de filtrage d'information sont développés. Le filtrage d'information (FI) est un processus dual à la recherche d'information comme le montre (Belkin et al., 1992). Il traite les documents provenant de sources sujettes à des modifications au cours du temps, et décide à la volée, si le document correspond ou pas aux besoins en information des utilisateurs, besoins modélisés au travers du concept de profils utilisateurs.

Les travaux de recherche effectués dans le filtrage d'information, prennent de plus en plus d'importance. On essaie d'adapter des techniques de recherche d'information au filtrage d'information. Bon nombre de systèmes de filtrage d'information sont basés sur des modèles de recherche d'information pour lesquels une fonction de décision, le plus souvent de type fonction seuil, est ajoutée. Filtrer un document revient d'une façon générale à mesurer une similitude entre un document et un profil, si le score est supérieur à un seuil, ce document est sélectionné sinon il est rejeté. Mais, l'absence de base de référence pour mesurer ce seuil et les pondérations adéquates associées aux profils et aux documents sont les problèmes majeurs rencontrés dans ce domaine. Cependant, plusieurs solutions sont proposées : une solution adaptative (Zhai et al., 1999) (Robertson et al., 1999a) (Boughanem et al., 2002) et une solution différée (Robertson et al., 1999b).

Ceci étant, toutes les solutions proposées dans le filtrage d'information ont le même objectif, sélectionner seulement des documents pertinents. Un problème majeur qui peut être soulevé et que ces systèmes demandent une masse d'information et force de calcul que les téléphones mobiles ne disposent pas.

## B)    RECHERCHE CONTEXTUELLE D'INFORMATION DANS UN ENVIRONNEMENT MOBILE

Les moteurs de recherche traditionnels considèrent peu le contexte de la recherche et ne sont pas adaptés à l'environnement mobile. Des travaux récents tentent d'améliorer les performances de recherche dans cet environnement. Une première catégorie de travaux a abordé les questions liées à l'adaptation de la recherche aux contraintes imposées par les fonctionnalités limitées des appareils mobiles. Des approches sont proposées pour adapter de la visualisation de la liste des résultats (Sweeney et al., 06) (Schofield et al., 02).

Une autre catégorie de travaux a porté sur l'exploitation du contexte de l'utilisateur mobile pour améliorer la précision des résultats de recherche. Dans (Yau et al., 03) les auteurs appliquent des techniques de data mining sur un historique d'usage, composé de données du contexte, d'actions et de données liées aux actions, pour construire les profils d'un utilisateur qui reflètent ses comportements, ses intérêts et ses intentions dans chaque situation.

Dans « Panayiotou et al., 06 » les auteurs exploitent l'importance du temps et de l'expérience (au travail, en vacances, etc) dans la personnalisation d'un portail de recherche de services web pour un utilisateur mobile. Ils proposent de construire un profil dynamique où les centres d'intérêt sont pondérés selon des zones temporelles apprises par l'étude de la routine journalière de l'utilisateur et ses activités dans chaque zone. De plus, pour modéliser le changement des préférences de l'utilisateur selon ses expériences, l'association des poids d'importance aux concepts du profil est établie pour chaque nouvelle expérience de l'utilisateur. Ces profils basés temps évoluent et sont maintenus sur la base des feedbacks utilisateurs.

Dans « Hattory et al., 07 » les auteurs traitent une méthode d'expansion basée sur la localisation. Leur méthode consiste a récupérer les coordonnées spatiales de l'utilisateur, les transformer en des mots contextuels (noms de places, d'activités liés à cette place) en utilisant un système d'information géographique et des techniques d'apprentissage des weblog, puis à pondérer ces mots contextuels sur la base de la comparaison de la probabilité globale du mot contextuel dans l'ensemble des documents du corpus cible et sa probabilité locale dans les documents retournés par la requête originale, en fin, la requête originale est étendue par les mots contextuels ayants les poids les plus élevés.

Dans « Ala-Siuru et al., 06 » les auteurs présentent une méthode qui combine l'identificateur de la cellule GSM à laquelle est connecté l'utilisateur et les adresses MACs des dispositifs Bluetooth à côté, pour caractériser une situation de l'utilisateur mobile. Un raisonnement par cas, peut être conduit alors, sur la base du contexte courant et des situations apprises, pour inférer le profil adéquat du mobile.

Nous remarquons que toutes ces techniques de recherche d'information contextuelle n'ont pas été étudiées dans un système de filtrage d'information.

En résumé nous pourrons dire que notre état de l'art va tourner autour des problématiques de recherche d'information contextuelle et du filtrage d'information.

## 4    DEMARCHE DE LA RECHERCHE

La démarche s'appuiera sur la construction d'un modèle de données mêlant filtrage d'information et Système d'Information dans le contexte de mobilité. Partant des connaissances théoriques et pratiques relevant de l'état de l'art, la formalisation proposée permettra de définir une spécification de système de filtrage permettant d'appréhender l'une ou plusieurs des différentes perspectives, qui pourront par la suite être analysées et employées pour répondre au besoin de l'utilisateur. Les travaux seront menés conformément à l'agenda présenté dans la figure ci-dessous.

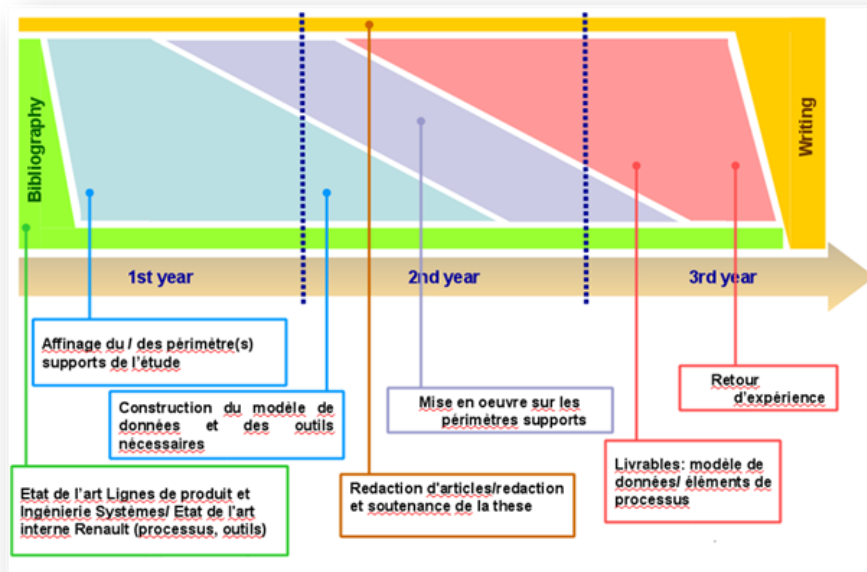

## 5   OBJECTIFS ET LIVRABLES

Il s'agit d'apporter des réponses aux problèmes identifiés.
Les livrables seront de plusieurs types :
•       Spécification d'un modèle de profilage adapté à un environnement mobile
•       Prototype d'un système de filtrage d'information
•       Retours d'expériences après mise en œuvre du système de filtrage sur une application réelle
•       Cahier des charges pour l'évolution des systèmes de filtrages dans un environnement mobile.